\documentclass[11pt]{article}  
  
\begin{document}

\title {Phase-transitions in a model for  the formation of herpes simplex
ulcers}

\author {Claudia P. Ferreira and J. F. Fontanari \\
Instituto de F\'{\i}sica de S\~ao Carlos \\
Universidade de S\~ao Paulo \\
Caixa Postal 369\\
13560-970 S\~ao Carlos SP, Brazil
\and
Rita M. Zorzenon dos Santos \\
Instituto de F\'{\i}sica \\
Universidade Federal Fluminense\\
Av. Litor\^anea s/n \\
24210-340 Niter\'oi RJ, Brazil} 
\date {} 
\maketitle

\begin{abstract} 

The critical properties of a cellular automaton model describing the
spreading of infection of the Herpes Simplex Virus in corneal tissue are
investigated through the dynamic Monte Carlo method. The model takes into
account different cell susceptibilities to the viral infection, as
suggested by experimental findings. In a two-dimensional square lattice
the sites are associated to two distinct types of cells, namely,
permissive and resistant to the infection. While a permissive cell becomes
infected in the presence of a single infected cell in its neighborhood,
a resistant cell needs to be surrounded by at least $R >1$ infected or
dead cells in order to become infected. The infection is followed by the
death of the cells resulting in ulcers whose forms may be dendritic
(self-limited clusters) or amoeboid (percolating clusters) depending on
the degree of resistance $R$ of the resistant cells as well as on the
density of permissive cells in the healthy tissue.  We show that a phase
transition between these two regimes occurs only for $R \geq 5$ and, in
addition, that the phase-transition is in the universality class of the
ordinary percolation. 

\end{abstract}

%\pacs{PACS: }
%\vskip2pc
\newpage
\small

\section{Introduction}\label{sec:level1}

One of the most common and intensively studied diseases among humans is
the herpes simplex virus (HSV) infection.  Apparently,
the unique symbiosis that exists
in nature between humans and the HSV allows the viral particles to remain
inactive (latent infection) in the cranial nerve ganglia after a primary
infection, producing frequently recurring localized infections during the
host's lifetime \cite{Liesegang}. The reactivation of HSV from latency may
occur at any time and it is characterized by active viral replication in
the epithelium causing vesicular eruptions in human mucosae and skin. The
rupture of these vesicles and the consequent cell necrosis leave the
characteristic herpetic lesion or ulcer.

Basically, there are two distinct types of herpes simplex virus, namely,
HSV Type I and HSV Type II. The former generally involves infection above
the waist (ocular and facial) while the latter infects tissues below the
waist. Here we discuss a mathematical model proposed to describe the
growth of corneal ulcers caused by HSV Type I \cite{Landini_94}.  This
infection is common and frequently causes corneal opacification.  
Traditionally the morphology of the corneal ulcers has been described as
either dendritic or amoeboid.  The dendritic ulcers are by far the more
frequent form and, though they are self-limited in general, ocasionally
they can enlarge progressively changing to the amoeboid form. This is
actually the natural course of the infection in the case of
immunocompromised hosts or of inappropriate use of topic corticosteroids.
In general, the amoeboid ulcers have a prolonged clinical course when compared to
the dendritic ones. Regardless of their morphology, the ulcers are
epithelial lesions that extend through the basement membrane whose swollen
epithelial borders contain active viral particles.

In order to
carry out  a more quantitative study of the ulcer morphology, the fractal
dimension of clinically diagnosed HSV ulcers (including both dendritic and
amoeboid forms) have been estimated suggesting that their outlines are
fractal objects \cite{Landini_92}. While the dendritic ulcers include
branching and linear lesions, the geographic ulcers are no longer linear and
as they increase in size, their perimeters become less and less
irregular. In addition to its usefulness as a classification tool, the
fractal properties of the ulcers  may give information on the underlying
mechanisms of viral spread within the epithelial
tissue. For instance, a theory based only on the neurotropism of HSV and
the dendritic-like distribution of nerve terminals can explain the
branching pattern observed in dendritic ulcers \cite{Dawson}, but it fails
to explain the decrease of the fractal dimension (perimeter irregularity)
with increasing ulcer sizes. An alternative explanation put forward
by Landini {\it et al} \cite{Landini_94}, which will be the
main focus of this paper, considers the ulcer shape as the natural outcome
of the contiguous spread of viral particles modulated by variations in the
cell susceptibilities to infection. To take
into account the fact that viruses only infect cells that have
appropriate receptor molecules on their surface, those authors
proposed a cellular automaton model for the HSV I spread in which the
corneal epithelial tissue is modeled by a two-dimensional lattice.  
In their model, each lattice 
site may be occupied either by a permissive cell (with probability $q$) or by a
resistant cell (with probability $1-q$). More pointedly,
a permissive cell becomes infected whenever there are at least one
infected cell in its neighborhood, while a resistant cell becomes
infected if the number of infected and dead neighboring cells is larger
than or equal to the integer parameter $R >1$ that measures the degree of
resistance of the cell \cite{Landini_94}.

The simulated ulcers obtained with the cellular automaton have the same
qualitative features of the clinical lesions and, in addition, for 
appropriate choices of the degree of resistance $R$ a dramatic change on
the morphology of the ulcers is observed as the initial concentration of
permissive cells $q$ increases beyond a certain value \cite{Landini_94}.  
This phenomenon was conjectured to be of a (qualitatively) similar nature
as the ordinary percolation phase transition. The main contribution of
this paper is to show, through the calculation of the dynamic and static
critical exponents, that in the cases where a phase-transition does occur
($R \geq 5$), the  transition belongs indeed to the universality
class of the ordinary percolation \cite{Stauffer,Isichenko}. To carry out
this analysis we use the so-called dynamic Monte Carlo method or spreading
analysis \cite{Grass-AP,Grass-MB} whose idea is to  study 
the spreading of the infection starting from a configuration with a 
single infected cell on the center of the lattice. 
Clearly this technique is very well suited to our
investigation since the characterization of the spreading behavior of the
infection is exactly the issue we address in this paper.

The remainder of the paper is organized as follows. Following Landini {\it
et al} \cite{Landini_94}, in Sec.\ \ref{sec:level2} we give the set of
rules that govern the evolution of the HSV I infection in a
two-dimensional square lattice and present the evidences for the existence
of a threshold phenomenon or phase transition for $R \geq 5$. In Sec.\
\ref{sec:level3} we characterize this phase transition using the dynamic
Monte Carlo method which allows the computation of the critical dynamic
exponents that describe quantitatively the spreading of the infection from
a single infected cell. Finally, some concluding remarks are presented in
Sec.\ \ref{sec:level5}.

%-----------------------------------------------------
\section{Model}\label{sec:level2}
%-----------------------------------------------------

The cellular automaton model is defined in a square lattice
consisting of $(L+1) \times (L+1)$ sites, where each site is associated to
a cell. Each cell is modeled by a four-state automaton corresponding to
the different states of this cell:  healthy permissive, healthy
resistant,
infected and dead. Except for the central cell, the initial state of any
cell in the lattice is set either as permissive or resistant with
probabilities $q$ and $1-q$, respectively, so that there are no dead cells
at the outset. The infection spreads from the single central
infected cell and the ulcer (i.e., the cluster of dead cells) grows
according to the following deterministic rules \cite{Landini_94}:

\begin{itemize} 
\item [(1)] An infected cell dies in the next time step. 
\item [(2)] A healthy permissive cell becomes infected if at least one of
its neighboring cells is infected.  
\item [(3)] A healthy resistant cell becomes infected
if at least $R>1$ of its  neighboring cells are infected or dead. 
\end{itemize} 
The neighborhood of a given
cell consists of its first and second nearest neighbors (Moore
neighborhood). 
The infection and subsequent death of a resistant cell surrounded by
$R$ or more dead cells is justified by the lack of tissue support. In
addition, this is necessary to prevent the occurrence of large ulcers with
small islands of resistant cells, which are not observed clinically
\cite{Landini_94}. The four-state automaton considered allows transitions of
one of the healthy states to the infected and dead states in a cyclic
manner. At each time step we perform a parallel updating of all cell
states.

For $R \leq 8$ we are dealing with a variant of the so-called
diffusion percolation process where the geometry changes via a dynamic
process and the nature of the growth depends on the local environment
\cite{Adler}. For finite lattice sizes and open boundary conditions the
above rules are repeated until either there are no more cells to infect or
an infected cell reaches the lattice boundary. These different modes of
termination generate dendritic (self-limited) and amoeboid (irrestricted)
ulcers, respectively. It is interesting to note that the ordinary
site percolation process is recovered for $R > 8$, since in this case a
resistant cell can never become infected and so the infection can
propagate only through the permissive cells.

\begin{figure}[tbp]
\vspace{6.5cm}
\includegraphics{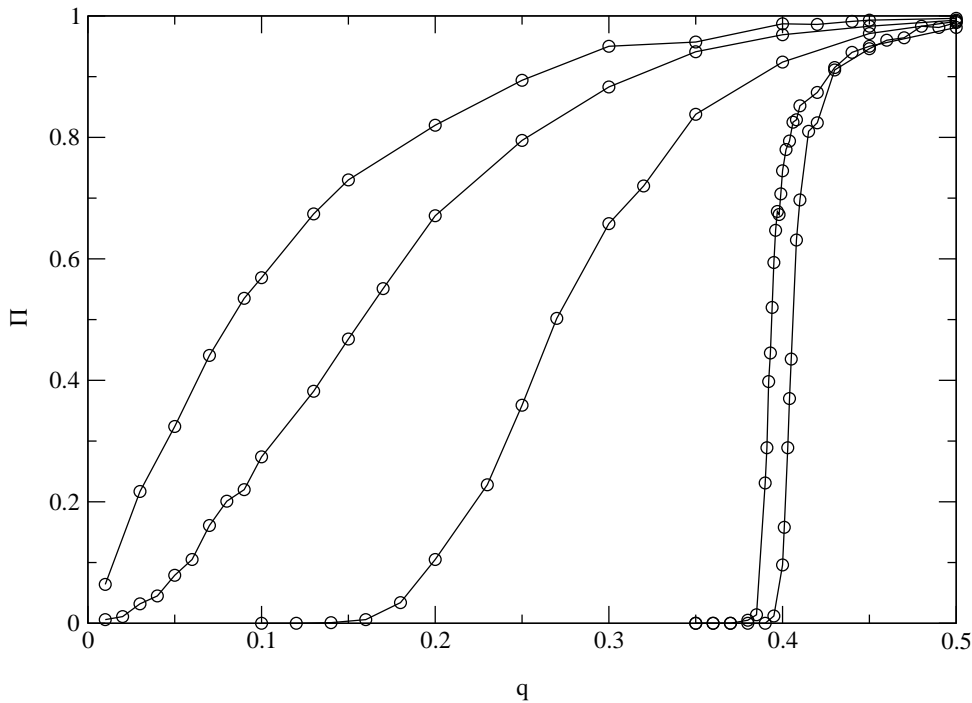} 
\vspace{2cm}
\caption{Percolation probability $\Pi$ as a function of the initial density 
of permissive cells $q$ for $L=1001$ and (left to right) $R=2,3,4,5$ and $6$.}
\label{fig:perc_R}
\vspace{1.cm}
\end{figure}

To illustrate the dependence of the different termination modes, and hence
ulcer forms, on the control parameters $R$ and $q$ of the model we present
in Fig.\ \ref{fig:perc_R} the fraction of irrestricted ulcers generated in
$1000$ runs. Each run corresponds to a different initial
configuration of the lattice. Clearly, this fraction can be identified with the
percolation probability $\Pi$ \cite{Stauffer,Isichenko}. Rather
interestingly, we have found that the results for $R \geq 6$ are
indistinguishable within the numerical precision. Actually, this is
expected since there is a preferred direction for the propagation of the
infection, namely, from the center to the lattice boundaries, and so only
the neighborhood facing the infection front matters to update the state of
a healthy cell. Since the largest size of the error bars in this as
well as in the next figure is twice the size of the symbols, they
were ommited for the sake of clarity. More importantly, we have found
that for $R \leq 4$, the results become independent of the lattice size
already for $L \geq 101$.

\begin{figure}[tbp]
\vspace{6.5cm}
\includegraphics{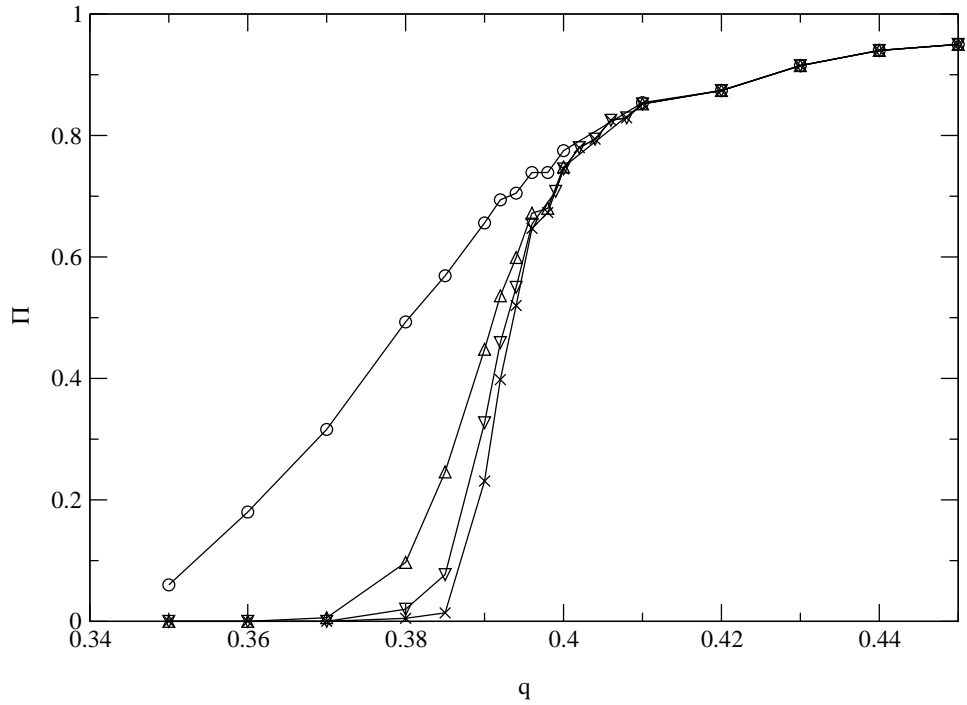} 
\vspace{2cm}
\caption{Percolation probability $\Pi$ as a function of the initial density 
of permissive cells $q$ for $R=5$  and $L=101 (\bigcirc), 401 (\bigtriangleup),
701 (\bigtriangledown)$ and $1001 (\times)$.}
\label{fig:perc_5}
\vspace{1.cm}
\end{figure}

However, for $R \geq 5$ the dependence on the lattice size, illustrated in
Fig.\ \ref{fig:perc_5} for $R=5$, indicates the occurrence of an atypical
threshold phenomenon at a critical value $q_c$ in the limit $L \to
\infty$. In fact, as $q$ increases from $0$ to $1$ the percolation
probability $\Pi$ vanishes for $q < q_ c$, undergoes a discontinuous
transition to some value $\Pi = \Pi_c > 0$ at $q=q_c$ and then increases
monotonically towards $1$. This transition is atypical in the sense that
$\Pi_c$ is not equal to $1$ above $q_c$, as in the case of the ordinary
percolation transition \cite{Stauffer,Isichenko}, which means that  in
this regime there is a finite probability that the infection
does not percolate, i.e., a dendritic ulcer is formed. The reason for that
is due simply to the fact that the spreading process starts from a single
central cell so that if the infection happens to percolate in a lattice of
a given size then it is certain to percolate in a smaller lattice too,
i.e., $\Pi (L_1) \geq \Pi (L_2)$ for $L_1 < L_2$. In particular, $\Pi (3)
= 1 - \left( 1 - p \right )^8$ yields an upper bound to $\Pi (\infty)$. Of
course, if the initial setting is such that there is an extensive number
of infected cells, say $\alpha L$ with $\alpha <1$, randomly distributed
over the bottom side of the lattice and periodic boundary conditions on
the lateral sides, then the usual result $\Pi_c=1$ is recovered
\cite{Levinshtein}. In fact, since the curves for different lattice sizes
do not cross, the standard finite size scaling analysis aiming at determining
both $q_c$ and the spatial correlation length
exponent $\nu_{\perp}$ for $R \geq 5$ (see, e.g., Ref.\ \cite{Stauffer}) fails
spectacularly and so we have to resort to other means to estimate those
quantities.

%

%----------------------------------------------------
\section{Spreading analysis}\label{sec:level3}
%----------------------------------------------------

We turn now to the analysis of the spreading behavior of the viral
infection starting from a single infected cell located in the center of a
lattice of infinite size. Finite size effects are absent because the
lattice size is taken large enough so that during the time we follow the
evolution the infection front can never reach the lattice boundaries.  
This of course sets an upper limit to the time we can follow the viral
spread and so, for instance, for lattices of size $L = 4005$ we let
the infection evolve up to $t = 2000$. As usual, we concentrate on the
time dependence of the following key quantities \cite{Grass-AP}: (i) the
average number of dead and infected cells $n(t)$; (ii) the survival
probability of the infection $p(t)$; and (iii) the average mean-square
distance over which the ulcer has spread $r^2(t)$. For each time $t$ we
carry out $10^4$ independent runs, hence $p(t)$ is simply the fraction of
runs for which there is at least one infected cell in the lattice at time
$t$.  At the transition point $q_c$ we expect that the measured quantities
obey the following scaling laws \cite{Grass-AP} 
\begin{eqnarray} p \left
( t \right ) & \sim & t^{-\delta} \label{delta} \\ n \left ( t \right ) &
\sim & t^{\eta} \label{eta}\\ r^2 \left ( t \right ) & \sim & t^{z}
\label{z} 
\end{eqnarray} 
where $\delta$, $\eta$ and $z$ are dynamic
exponents. Since the fractal dimension $d_f$ of the ulcer at a given time
$t$ is defined as $n (t) \sim r^{d_f}$ we have
\begin{equation}\label{df} 
d_f = 2 \frac{\eta}{z} \end{equation} 
at the critical point. Note that this equation is different from the one used in
the studies of directed percolation (see, e.g., Ref.\ \cite{Grass-JPA})
because in the present case all runs generate an ulcer and so $n(t)$ as
well as $r^2(t)$ are averages taken over all runs.

\begin{figure}[tbp]
\vspace{6.5cm}
\includegraphics{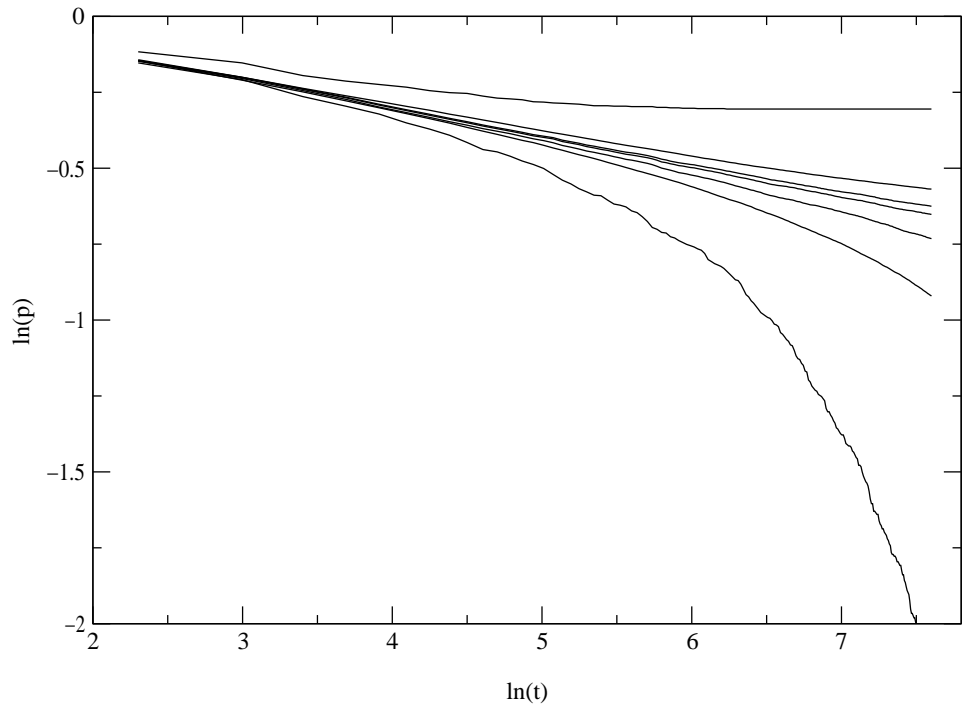} 
\vspace{2cm}
\caption{The log-log plot of $p(t)$ as a function of $t$
for $R=5$ and (top to bottom) $ q=0.4,0.395,0.3947,0.3945,0.394,0.393$
and $0.39$.}
\label{fig:logP}
\vspace{1.cm}
\end{figure}

\begin{figure}[tbp]
\vspace{6.5cm}
\includegraphics{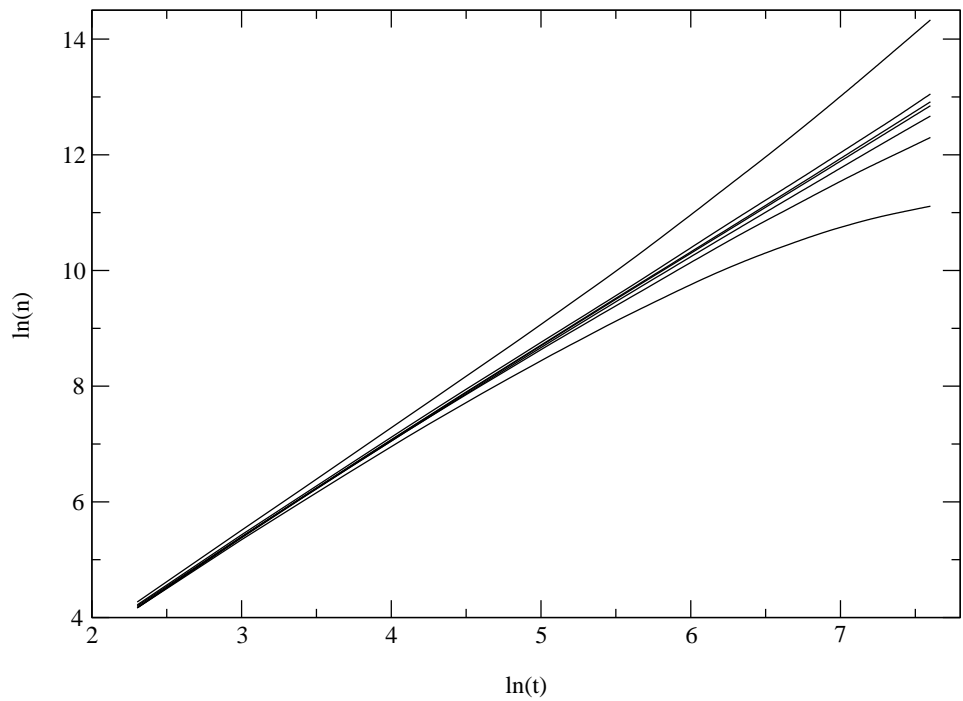} 
\vspace{2cm}
\caption{Same as fig. \ref{fig:logP} but for $n(t)$.}
\label{fig:logn}
\vspace{1.cm}
\end{figure}

\begin{figure}[tbp]
\vspace{6.5cm}
\includegraphics{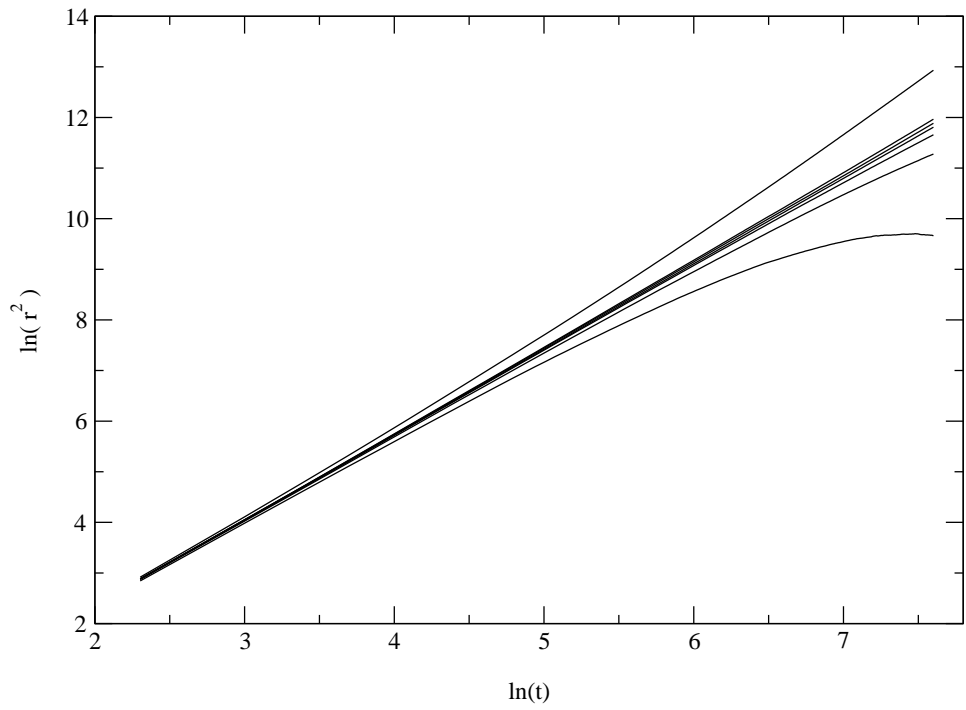} 
\vspace{2cm}
\caption{Same as fig. \ref{fig:logR} but for $r^2(t)$.}
\label{fig:logR}
\vspace{1.cm}
\end{figure}

In Figs. \ref{fig:logP}, \ref{fig:logn} and \ref{fig:logR} we present
log-log plots of $p(t)$, $n(t)$ and $r^2(t)$, respectively, as functions
of $t$ in the vicinity of the critical point for $R=5$.  The asymptotic
straight lines observed in these figures are the signature of critical
behavior while upward and downward deviations indicate supercritical ($q >
q_c$) and subcritical ($q < q_c$) behaviors, respectively. We recall
that in the subcritical regime only dendritic ulcers are formed, 
while in the supercritical regime the formation
of amoeboid ulcers is much more frequent (see Fig.\ \ref{fig:perc_5}). 
The data shown in Fig.\ \ref{fig:logP} yield $q_c = 0.3945 \pm 
0.0002$ where the error is estimated by determining two values of $q$ as
close as possible to the critical point for which upward and downward
deviations can be observed. A precise estimate for the dynamic critical
exponents is obtained by considering the local slopes of the curves shown
in the previous figures. For instance, the local slope $\delta (t)$ is
defined by \cite{Grass-JPA} 
\begin{equation} 
-\delta \left ( t \right ) = \frac{\ln \left[ p(t)/p(t/8) \right]}{\ln 8} , 
\end{equation}
which for large $t$ behaves as \begin{equation} \delta \left ( t \right )
\sim \delta + \frac{a}{t} \end{equation} where $a$ is a constant.  
Analogous expressions hold for $\eta(t)$ and $z(t)$. Hence plots of the
local slopes as functions of $1/t$ allow the calculation of the critical
exponents. Applying this procedure for the critical curves we find the
exponents $\delta = 0.0870 \pm 0.0001 $, $\eta = 1.5866 \pm 0.0007 $ and
$z = 1.6843 \pm 0.0003$. The errors in the critical exponents are, as
usual, the statistical errors obtained by fitting the local slopes by
straight lines in the large $t$ regime. We expect, of course, that the
(uncontrolled) systematic errors are much larger than those. Using Eq.\
(\ref{df}) we obtain $d_f = 1.8840 \pm 0.0005$ which is in very good
agreement with the analytical prediction for the ordinary percolation $d_f
= 91/48 \approx 1.896$ \cite{Stauffer,Isichenko}.

\begin{figure}[tbp]
\vspace{6.5cm}
\includegraphics{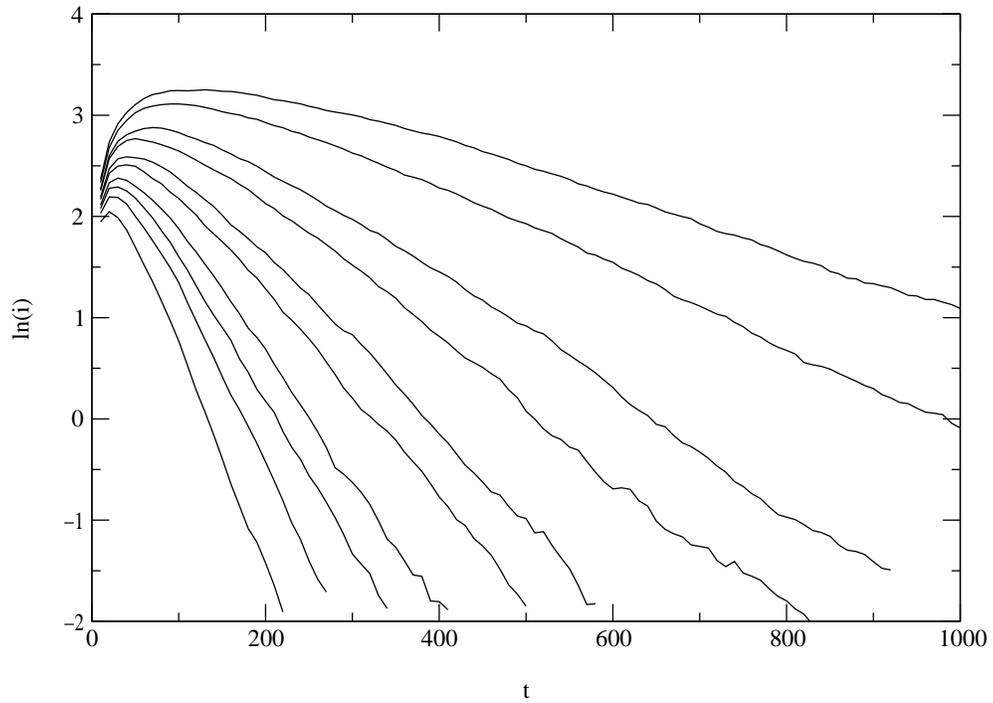} 
\vspace{2cm}
\caption{The log-linear plot of $i(t)$ against $t$
for $R=5$ and (top to bottom) $ q_c - q = 0.009, 0.011, 0.014, 0.016, 0.019, 0.021, 
0.024, 0.027, 0.029$
and $0.034$. }
\label{fig:logI}
\vspace{1.cm}
\end{figure}

\begin{figure}[tbp]
\vspace{6.5cm}
\includegraphics{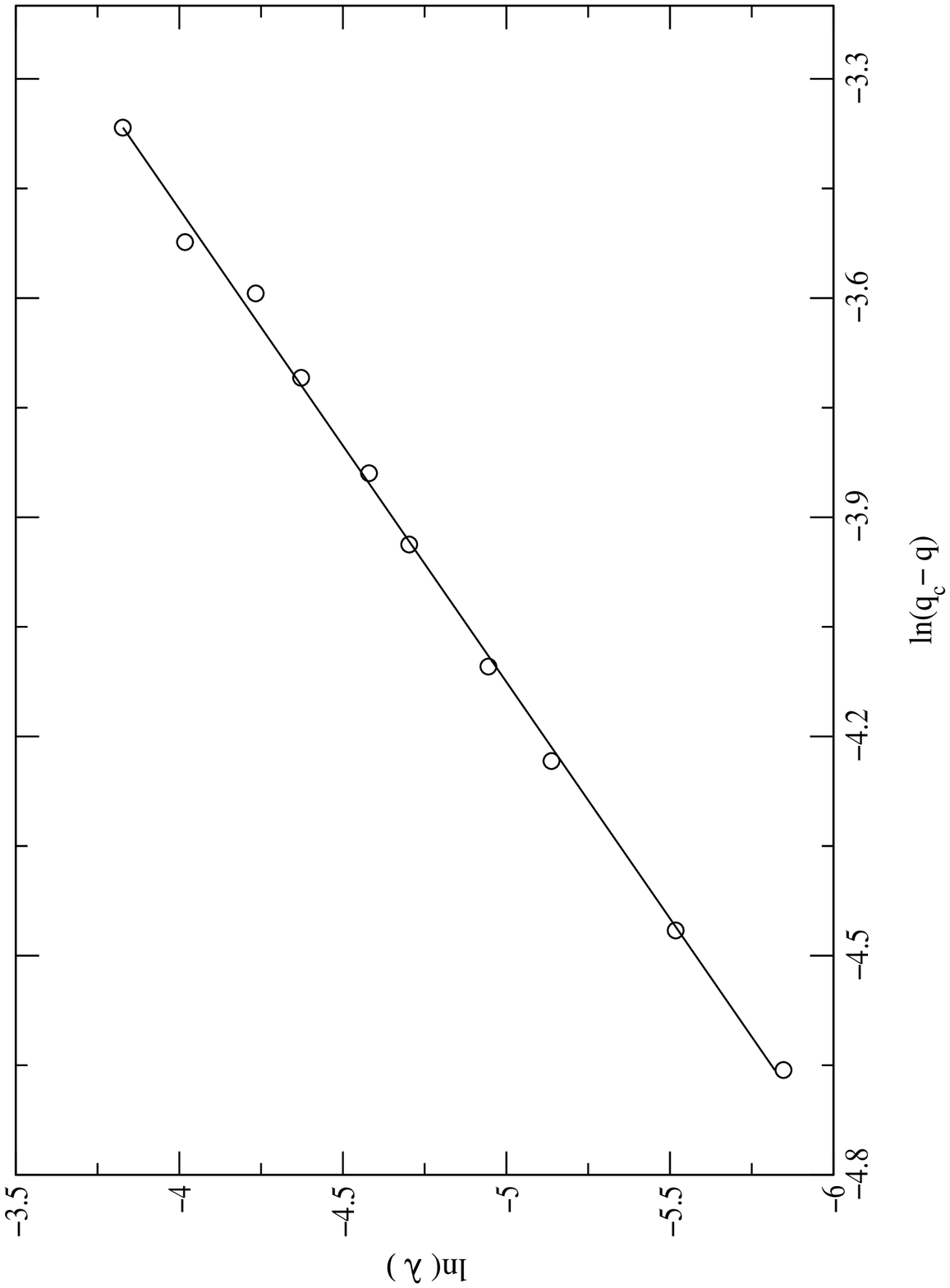} 
\vspace{2cm}
\caption{The log-log plot of the time decay constant $\lambda$ against $ q_c - q $
for $R=5$. The slope of the straight line yields $ \nu_{\parallel} \approx 1.54 \pm 0.03$.}
\label{fig:logL}
\vspace{1.cm}
\end{figure}

As the dynamic exponents $\delta$, $\eta$ and $z$ for the ordinary
percolation problem are not very well known, to show unambiguously that
this ulcer formation model belongs to the universality class of the
ordinary percolation we ought to estimate the static exponents $\beta$ and
$\nu_{\perp}$.  We recall that the exponent $\beta$ gives a measure of how
the fraction of lattice cells belonging to an infinite cluster vanishes as
the percolation threshold $q_c$ is approached in the supercritical regime
while $\nu_{\perp}$ is the correlation-length exponent in the space
direction. To do so we calculate first the exponent $\nu_{\parallel}$
which governs the decay of the concentration of infected cells $i(t)$ in
the subcritical regime. In fact, since in this regime the correlations are
short-ranged one expects $i(t)$ to decay exponentially \cite{Grass-AP} 
\begin{equation}\label{i_decay} i(t) \approx ~ A \left ( q \right )  \exp
\left [ - \left ( q_c - q \right )^{\nu_{\parallel}} t \right ] \ \ \ \ \
t \rightarrow \infty \end{equation} 
where $A \left ( q \right )$ is some
time independent function.  Fig.\ \ref{fig:logI} not only illustrates the
adequacy of this assumption but permits also the evaluation of the decay
constant 
\begin{equation}\label{lambda} \lambda = \left ( q_c - q \right
)^{\nu_{\parallel}} 
\end{equation} 
from the asymptotic slopes of the
curves $\ln i$ vs. $t$. The results presented in Fig.\ \ref{fig:logL},
showing the dependence of $\lambda$ on the distance $q_c - q$ from the
critical point, allows the calculation of the exponent $\nu_{\parallel}$
as the slope of the straight line, yielding $ \nu_{\parallel} = 1.54 \pm
0.03$. Once this exponent is known we can use the scaling relations
$\beta=\nu_{\parallel} \ \delta$ and $\nu_{\perp} = z \nu_{\parallel}/ 2$
\cite{Grass-AP} to estimate the static exponents. We find $\beta = 0.134
\pm 0.003$ and $\nu_{\perp} = 1.30 \pm 0.03$ which, within error bars, are
in agreement with the exact values of the corresponding exponents of the
ordinary percolation, namely, $\beta = 5/36 \approx 0.139$ and
$\nu_{\perp} = 4/3 \approx 1.333$ \cite{Stauffer,Isichenko}.


We have carried out a similar spreading analysis for $R \geq 6$ and, as
hinted in Fig.\ \ref{fig:perc_R}, we have found a slightly larger
percolation threshold, namely, $q_c = 0.4075 \pm 0.0002$ which, within
error bars, is shown to be independent of the value of $R \geq 6$.
Furthermore, since the larger the resistance parameter $R$, the more
similar the ulcer formation problem is to the ordinary site percolation,
we have found the same dynamic and static critical exponents as for the
case $R=5$, as expected.

%----------------------------------------------------
\section{Conclusion}\label{sec:level5}
%----------------------------------------------------

Using the dynamic Monte Carlo method  we have shown
unambiguously that the phase transition observed in the model for
formation of herpes simplex ulcers proposed by Landini {\it et al}
\cite{Landini_94} belongs to the universality class of the ordinary
percolation.  The value of this finding should not be underrated since the
infection process actually resembles a diffusion percolation process where
the growth depends on the local environment, in the sense that the
decision on whether or not a resistant cell will become infected depends
on the time-dependent states of several of its neighbors. Furthermore,
since the ulcer formation model described here may be thought of as a damage 
spreading process, one could expect that the transition were in 
the universality class of the $(2+1)$ directed percolation instead.
However, as pointed out by Grassberger \cite{Grass-JSP_0}, this is not so because
in the ulcer formation  model the damage never heals (even if it does not spread), i.e.,
the probability that an infected or dead cell becomes healthy is zero.

The finding that for $R <5$ the model does not present a phase transition
reflects the nontrivial role played by the resistance parameter $R$ in
this percolation process.  In these noncritical cases the probability that
an infinite or irrestricted ulcer is generated is given by the smooth
size-independent curves shown in Fig.\ \ref{fig:perc_R}. Similarly to a
noncritical forest fire model \cite{Bak,Grass-JSP} the growing of this
type of ulcer may be characterized by infection fronts with fractal
dimension $D$ whose value probably depends on the resistance parameter $R$ (of
course, $D=1$ for $R=1$). An additional feature that makes the
quantitative study of this viral spreading model rather challenging is the
result that the percolation probability curves for different lattice sizes
do not cross (see Fig.\ \ref{fig:perc_5}), which complicates enormously
the estimate of the percolation threshold and critical exponents through
the standard finite size scaling method.

Some remarks on the biological interpretation of our results are in order.
According to the specialized literature \cite{Liesegang,Landini_94,Landini_92,Dawson},
amoeboid ulcers are, in general, observed in immunocompromised patients or
in patients that made inappropriated use of corticosteroids.
In the present model, these conditions would correspond to a decrease of the
degree of resistance $R$ of the resistant cells or to an increase of
the initial concentration $q$ of permisssive cells. Although this model
does not take into account the recurrent characteristic of this kind of 
infection, in which case the variability of $q$ would probably play an 
important role, nor the possibility of variation of $R$ during the course
of the infection, its predictions 
are in qualitative agreement with the clinical observations. In fact,
Fig.\ \ref{fig:perc_R}  points out the prevalence of amoeboid ulcers
when $R$ decreases or $q$ increases. This agreement lends support to
the hypothesis that the morphology of the ulcers is determined
by the viral spreading through 
cells with different susceptibilities to infection. 

To conclude we should mention that an extension of the original model
proposed by Landini {\it et al} in which both the regeneration of dead
cells as well as the spontaneous outbreak of infection anywhere in the
lattice are taken into account has already been considered in the
literature \cite{Coutinho}. Interestingly, in this case the viral
spreading model becomes very similar to the critical forest fire model
with immune trees \cite{Drossel,Albano}. In particular, the resistance
parameter $R$ of the ulcer formation model is akin to the immunity
probability, i.e., the probability that a tree is not ignited though one
of its neighbors is burning. According to a conjecture
put forward by Grassberger \cite{Grass-JSP_0}, the extended ulcer
formation model should be in the universality class of
directed percolation, since it allows for the regeneration
of dead cells. This suggestion is strengthened by the finding that  the
forest fire model with immune trees is in that class of 
universality \cite{Albano}.

\bigskip

\bigskip

The work of J.F.F. is supported in part by Conselho
Nacional de Desenvolvimento Cient\'{\i}fico e Tecnol\'ogico (CNPq)
and  Funda\c{c}\~ao de Amparo \`a Pesquisa do Estado de S\~ao Paulo 
(FAPESP), Proj.\ No.\ 99/09644-9.
The work of R.M.Z.S. is partially supported  by CNPq.
C.P.F.  is supported by FAPESP. We thank FAPESP for supporting 
R.M.Z.S's visit to S\~ao Carlos where part of her work was done.

%\begin{references}

\end{document}